\documentclass[a4paper,fleqn]{cas-dc}

\usepackage[authoryear]{natbib}
\begin{document}
\let\WriteBookmarks\relax
\def\floatpagepagefraction{1}
\def\textpagefraction{.001}
\shorttitle{Pronunciation generation for foreign language words}
\shortauthors{Wei Wang et~al.}

\title [mode = title]{Pronunciation Generation for Foreign Language Words in Intra-Sentential Code-Switching Speech  Recognition}                      
\author[1,2]{Wei Wang}[style=chinese,orcid=0000-0002-1765-0486]
\ead{wei.wang@imsl.org.cn}

% \author[2]{Xiao Song}[style=chinese]
% \ead{xiao.song@imsl.org.cn}

\author[1]{Chao Zhang}[style=chinese]
\ead{14042@ahu.edu.cn}

\author[1]{Xiaopei Wu}[style=chinese]
\cormark[2]
\ead{wxp2001@ahu.edu.cn}

% \author[2,3]{Xin Wang}[style=chinese]
% \cormark[1]
% \ead{xin.wang@imsl.org.cn}

\address[1]{Anhui Provincial Key Laboratory of Multimodal Cognitive Computation, School of Computer Science and Technology, Anhui University, Hefei 230601, China}
% \address[2]{Peking University Shenzhen Institute, Shenzhen 518000, China}
% \address[3]{Shenzhen Raixun Information Technology Co., Ltd., Shenzhen 518000, China}

% \cortext[cor1]{Corresponding author}
\cortext[cor2]{Principal corresponding author}

\begin{abstract}
Code-Switching refers to the phenomenon of switching languages within a sentence or discourse. However, limited code-switching , different language phoneme-sets and high rebuilding costs throw a challenge to make the specialized acoustic model for code-switching speech recognition. In this paper, we make use of limited code-switching data as driving materials and explore a shortcut to quickly develop intra-sentential code-switching recognition skill  on the commissioned native language acoustic model, where we propose a data-driven method to make the seed lexicon which is used to train grapheme-to-phoneme model to predict mapping pronunciations for foreign language word in code-switching sentences. The core work of the data-driven technology in this paper consists of a phonetic decoding method and different selection methods. And for imbalanced word-level driving materials problem, we have an internal assistance inspiration that learning the good pronunciation rules in the words that possess sufficient materials using the grapheme-to-phoneme model to help the scarce. Our experiments show that the Mixed Error Rate in intra-sentential Chinese-English code-switching recognition reduced from 29.15\%, acquired on the pure Chinese recognizer, to 12.13\% by adding foreign language words' pronunciation through our data-driven approach, and finally get the best result 11.14\% with the combination of different selection methods and internal assistance tactic.
\end{abstract}


\maketitle
\hbadness=10000
\vbadness=10000
\tolerance=10000
\hfuzz=150pt

\section{Introduction}
\label{sec1}
%% background %%
    Code-Switching (CS) is a common oral phenomenon for many multilingual speakers that different languages coexist in sentences. As \citep{Sankoff1981} defined, CS can be categorized into inter-sentential switching and intra-sentential switching. In the intra-sentential case we focus on in this paper, foreign words usually appeared in a native language sentence as loan words. Compared with monolingual automatic speech recognition (ASR) systems, the hindrances in the CS recognition are summarized as follows: (1) the lack of CS training data, (2) the phonemes difference in different languages, (3) the  accent issues on foreign languages spoken by native language speakers.

%% mainstream method %%
    In recent years, the related works in tackling CS have been continuously deepened. Some mainstream methods focus on building the mixed-language acoustic model (AM) and the language model (LM). Due to the limited CS speech data, the phoneme-sharing method in multiple languages has been applied broadly to decrease the size of mixed-language AM units, where the sharing methods can be divided into phoneme-merging \citep{Lin2009,Li2011,Sivasankaran2019} and using a universal phoneme set such as IPA \citep{Smith2000} generally, however, those may increase the risk of inter-language substitution errors due to ruining the context of some triphones in each language. Some multi-task-learning (MTL) technologies \citep{Huang2013,Yilmaz2016} also have been explored in CS, where the recognition for accented speech \citep{Mendes2019} and the recognition around the switching  position \citep{Chen2016b} have been improved by transferring knowledge between different tasks.
    Generally, in the intra-sentential CS case, the main part of speech is still native language (NL) and foreign words occupy less often, therefore, under the commissioned native language acoustic mode (NL-AM) that shows good recognition and robustness in the real scenario since it has abundant training data, it is a valuable work to explore a shortcut to preserve its stability in native language and extend its capability to foreign words. 

%% proposed method %%
    For intra-sentential CS speech recognition, this paper proposes a data-driven scheme to generate mapping pronunciations for foreign words to meet intra-sentential CS speech recognition, concretely, the reliable pronunciations are given by the Grapheme-to-Phoneme (G2P) model which was trained on the seed lexicon, where  making this seed lexicon is a core data-driven work and this paper adopts phonetic decoding for candidate generation and average posterior estimation (APE) or phoneme Confusion Network (PCN) for candidate selection. 
    Compared with the seed lexicon made by \citep{Huang2019}, this paper proposes the following improvements: (1) A purely data-driven approach without any linguist pronunciation labeling. (2) In acoustic-based candidate selection, we adopt the average utterance-level posterior probability of candidate to give an acoustic score. (3) For pronunciation prediction, we use the popular transformer-based \citep{Zhang2017} sequence-to-sequence (seq2seq) architecture in natural language processing (NLP) field to design our G2P model.

%% improved method %%
    However, foreign words' different occurrences in the CS corpus lead to imbalanced word-level driving materials in the data-driven approach, which further cause imbalanced driving processing, and the pronunciation qualities with fewer materials always show poor performance. In this work, for the imbalanced driving material issue, we further propose an internal assistance strategy between the words with sufficient materials and the words with scarce materials to improve the pronunciation qualities in the seed lexicon overall.

%% section spliting %%
    The rest of this paper is organized as follows. 
    Section 2 is a summary
    of related works on retraining-free and phonetic decoding methods for CS. 
    Section 3 describes candidate generation and selection methods for seed word pronunciation. 
    Section 4 introduces the pronunciation prediction works that contain building seed lexicon, describing the architecture of the transformer G2P model and proposing a novel internal assistance method to improve the seed lexicon.
    Section 5 gives the detailed experimental configuration and result. 
    Section 6 concludes the advantages of the proposed methods and lists some valuable future works.

\section{Related Works}
\label{sec2}
%% pronunciation mapping method %%
    Compared to the available data in monolingual ASR, the CS corpus is very limited \citep{Ganji2019,Lyu2010,Li2012,Shen2011,Chan2005,Lyu2006}, hence some CS recognizers adopt the retraining-free methods that expand new language recognition capabilities in existing monolingual ASR system instead of rebuilding mixed-language AM/LM.
    \citep{Yu2009} proposed and compared four approaches for CS recognition under the constraint of native language acoustic model (NL-AM) in real-time, in that case, the foreign words were expressed in the native language phonemes set through phoneme/senone mapping using the least Kullback-Leibler Divergence, and achieved the best result among the AM merging techniques.
    Base on the NL-AM, Pronunciations Generation in foreign words is considered as another low-cost solution for intra-sentential CS speech recognition, so the core work will focus on generating good mapping pronunciations to foreign words in native language phonemes, which work is similar to automatic lexicon learning \citep{McGraw2013,Lu2013,Chen2016,Tsujioka2016,Zhang2017} in solving the out-of-vocabulary issues.
    \citep{Laurent2010} proposed acoustic-based phonetic decoding and iterative filtering methods for proper nouns' phonetic transcription. 
    \citep{Bhuvanagiri2010} and \citep{Bhuvanagirir2012} built a Hindi-English ASR system based on the existing monolingual AM using the mapped lexicon and the modified language model. 
    \citep{Modipa2013} constructed the Sepedi-English ASR system based on a Sepedi speech decoder, where the pronunciations of English words were obtained from the Sepedi language phonetic decoder and then added into the original lexicon.
    \citep{Huang2019} obtained high-quality foreign words' pronunciations from a grapheme-to-phoneme (G2P) model trained on linguist/data-driven lexicon, where the data-driven method consisted of a phonetic decoding on foreign words spoken by native language speakers  generating method and a rover-like \citep{Fiscus1997} Phoneme-Confusion-Networks with acoustic score ranking method. 
    Also, in building mixed-language AM for the Mandarin-English CS task, \citep{Guo2018} adopted phonetic decoding method to correct mismatched pronunciations by decoding .

\section{Pronunciation Generation For Seed Word }
\label{sec3}
%% summary %%
    This section mainly introduces the data-driven way to obtain good pronunciations of foreign seed words.
	Approaches in generating pronunciations  can be divided into manual and data-driven categories. On the manual side, people will consider the rationality on pronouncing perception, but that may be time-consuming and imprecise due to accent problems which are often neglected. On the data-driven side, we will use phonetic decoding technology where the decoding results are consistent with the similarity in acoustics and take native accents into account.
	
	The phonetic decoding method is a data-driven way to obtain foreign words' pronunciations, that is, the NL-AM based phonetic decoder is used to decode the foreign words' audio segments to obtain the native phoneme sequence as candidate pronunciations. The source to obtain foreign words' audio segment is mainly derived from the speech data of native or foreign speakers, due to the CS speech recognition task is oriented to native speakers, in this article we use the audio segments in a limited CS corpus spoken by native speakers as driving materials.

%% 3.1 extracting audio segments %%
\subsection{Extracting Audio Segments}
\label{sec3.1}
     For foreign words' audio segmentation, we need to first obtain their start/end timestamps. The timestamp acquisition is achieved by speech-text forced alignment on the AM, and the  general method is to build mixed-language GMM-HMM based AM with a combination of  native language lexicon and foreign language lexicon.  
     \citep{Huang2019} maintain the retraining-free way that they used foreigner’s audio segments as driving materials to obtain mapped lexicon by phonetic decoding to execute forced-alignment on the NL-AM, though this method avoided pre-training a mixed-language AM, but foreign words' pronunciations with foreign accent may cause inaccurate alignments. In this work, we still choose to pre-train a mixed-language GMM-HMM based AM for alignment and segmentation of foreign words.

%% 3.2 phonetic decoding %%
\subsection{Phonetic Decoding}
\label{sec3.2}
    Different from the word-level ASR system, a phonetic decoder is built on the decoding graph with a phoneme-level LM. Based on the NL-AM, we build  a phonetic decoder $\mathbf{P}$ , and then decode the foreign words' audio segments in high acoustic weight setting to obtain the native language phoneme sequences with high acoustic similarity.
    
%% segments %%
    For a foreign language word $w$, we extract its embedded utterances subset $\textbf{O}_{w} = \{O_{1},O_{2},\cdots,O_{M_{w}}\}$ in the limited CS corpus, where $M_{w}$ denotes the number of utterances which contain the word $w$. As \autoref{sec3.1} introduces,  we extract its segments set ${S}_{w}$ in $\textbf{O}_{w}$ through forced-alignment in mixed-language GMM-HMM:	
	\begin{equation}
	{S}_{w} = \{s_{i}\mid i=1,2,\cdots,k\}
	\label{eq1}
	\end{equation}
    where ${S}_{w}$ collects $k$ segments from \textbf{O}$_{w}$ ($k \ge M_{w}$, because there is at least one per utterance), also the $k$ means the driving volume of $w$ in the data-driven approach.
%% decoding %%
	Then each segment ${s}$ in ${S}_{w}$ will be fed into phonetic decoder $\mathbf{P}$ and get the $n$ best results:
	\begin{equation}
	\mathbf{P}(s) = \{p_1,p_2,\cdots,p_n\}
	\label{eq2}
	\end{equation}
%% candidates %%
	Finally, we will merge all decoding results into an initial candidates set $\Phi_{w}$ for $w$:
	\begin{equation}
	\Phi_{w} = \mathbf{P}(s_{1})\cup \mathbf{P}(s_{2})\cup \cdots \cup  \mathbf{P}(s_{k})
	\label{eq3}
	\end{equation}

%% 3.3 Average Posterior Estimation  %%
\subsection{Candidate Selection}
	From \autoref{sec3.2}, we utilize phonetic decoding approach as the source for producing candidate pronunciations to foreign words, but those decoded candidates may include a lot of noises due to the acoustic difference in different languages and the possible inaccurate segments-cutting. Hence, we need to filter and select from these raw candidates to obtain high-quality candidates which can bring good recognition results,  here we will introduce some selection methods in this section.

%% 3.3.1 Average Posterior Estimation  %%
\subsubsection{Average Posterior Estimation}
\label{sec3.3.1}
\begin{figure*}[pos=htbp,width=0.9\textwidth,align=\centering]
        \centering
        \includegraphics{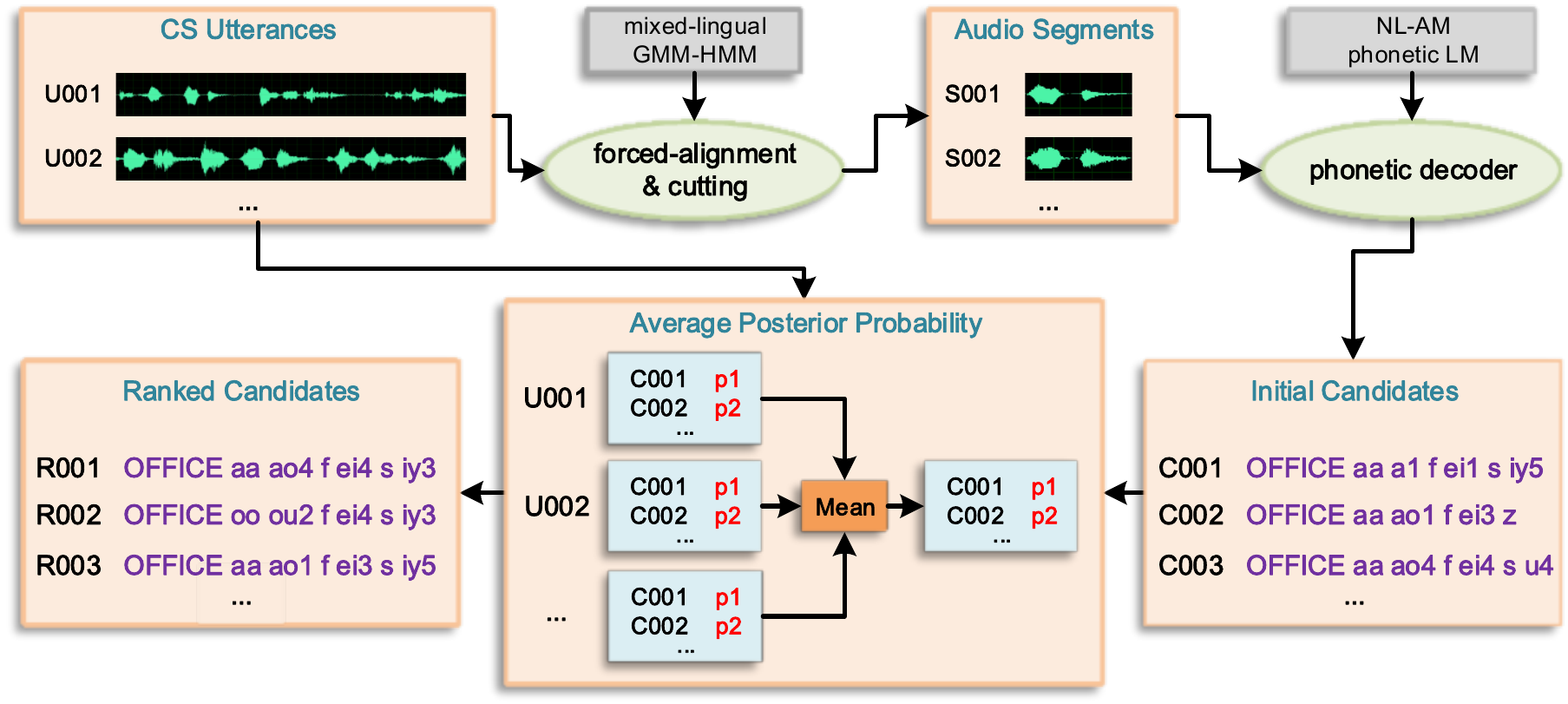}
        \caption{Average posterior estimation on the candidates of the foreign word \textsl{office} in Chinese-English CS.}
        \label{ape}
    \end{figure*}

     In data-likelihood-reduction based pronunciation selection criterion, \citep{Zhang2017} collected the acoustic evidence  using conditional data likelihood of utterance $O_{u}$ given the pronunciation of word $w$ being candidate $b$: $\tau_{u,w}^{b} = P(O_{u}|w,b)$, then they replaced  $\tau_{u,w}^{b}$ with the posterior statistic $\gamma_{u,w}^{b} = P(w,b|O_{u})$ by Bayes’ rule and further removed low-quality candidates by comparing their average utterance-level posterior statistics.
      Similarly, we evaluate the overall acceptance in original utterances by calculating the average utterance-level posterior probability to each candidate with the following formula:
	\begin{equation}
	\gamma_{w}^p = \frac{1}{M_{w}}\sum_{u=1}^{M_{w}}P(w,p\mid O_{u})
	\label{eq4}
	\end{equation}
	where $P(w,p\mid O_{u})$ represents the posterior probability of $w$ pronouncing $p$ in the utterance $O_{u}$ \footnote{If $w$ appears multiple times in $O_{u}$, we will calculate its mean of posterior probabilities}.  
%% ranking %%
	According to the comparison of $\gamma_{w}^p$, we can get the ranked table in initial candidates $\Phi_{w}$ and select the best results subsequently. A pronunciation selecting example to the foreign word \textsl{office} in  Chinese-English CS is showed in \autoref{ape}.

%% 3.3.2 Phoneme Confusion Network  %%
\subsubsection{Phoneme Confusion Network}
\label{sec3.3.2}
    Another novel approach in selecting candidates is to combine the best results on initial candidates. \citep{Huang2019} proposed to build a phoneme confusion network (PCN) on the candidates of a particular word to combine the most generalized candidates (voting-based), where building PCN is a ROVER-like \citep{Fiscus1997}  method that all candidate sequences are combined into a single, minimal cost word transition network via iterative applications of dynamic programming alignments. For example, we get 10 candidates of the word \textsl{health} from phonetic decoder:
    \begin{enumerate}
    \itemsep=0pt
        \item h ai2 ii iu5 x i3
        \item h ai2 ii iu5
        \item h ai2 ii iao1 x i2
        \item h ai2 ii iao4
        \item h ai2 ii iao1 x i4
        \item h ai2 ii iao1 x i1
        \item h ai2 ii iao1 x i3
        \item h ai2 ii iao2 s iy3
        \item h ai2 ii iao3 s iy3
        \item h ai2 ii iao4 s iy3
     \label{candidates of pcn}
    \end{enumerate}
    
	\begin{figure}[ht]
    	\centering
    	\includegraphics[width=\linewidth]{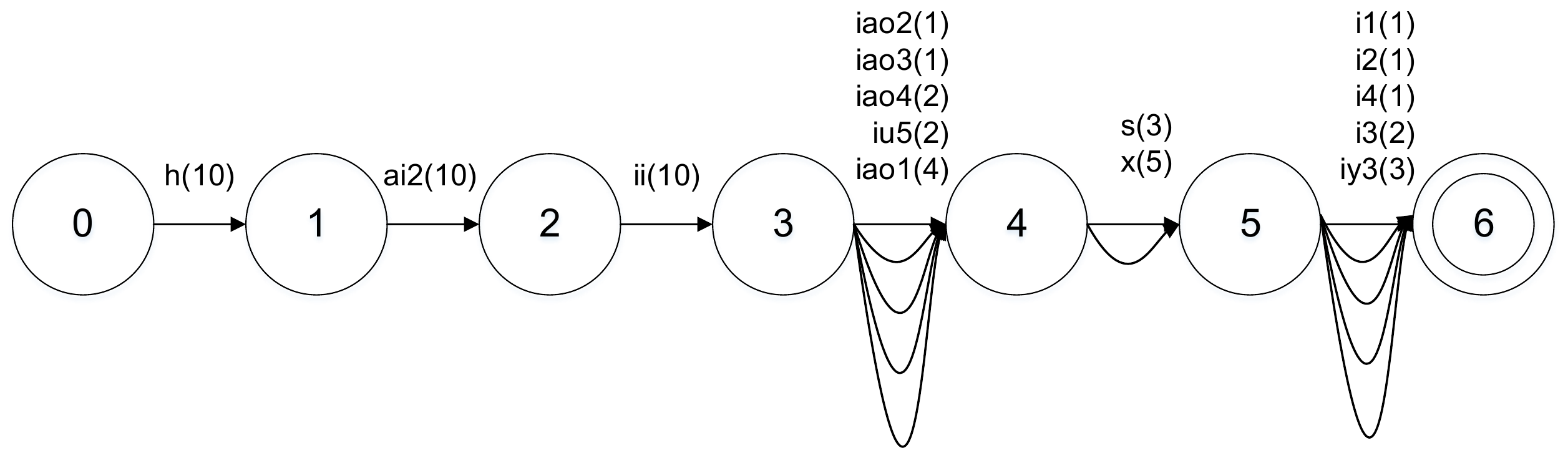}
    	\caption{Phoneme confusion network built with 10 candidates for word 'health'}
    	\label{pcn}
	\end{figure}
    
    then the corresponding phoneme confusion network is built as \autoref{pcn} shows, where the numbers in parentheses represent the number of times a phoneme appears and we can summary 4 best phoneme sequences (with most votes) for 'health':
    \begin{enumerate}
    \itemsep=0pt
        \item h ai2 ii iao1 x iy3
        \item h ai2 ii iao1 x i3
        \item h ai2 ii iao1 s iy3
        \item h ai2 ii iao1 x i1
     \label{results from pcn}
    \end{enumerate}
    It is worth noting that the summary results from PCN may produce new variants such as the result 'h ai2 ii iao1 x iy3' which has not appeared in original candidates.

\section{Pronunciation Prediction}
\label{sec4}
%% summary %%
    For various CS scenarios, we want to achieve the pronunciations of any foreign word (including out-of-set foreign words), but our available pronunciations come from the closed set $F$ composed of all foreign words in the CS corpus.
    Grapheme-to-Phoneme \citep{Rugchatjaroen2019,Bisani2008,Bellegarda2005} approach is the process of converting the written form  to the pronunciation of a word and is  a seq2seq method from character sequence to phoneme sequence essentially.
    In this paper we use a G2P model to learn the pronunciation rules in the best candidates of closed set $F$, that is, we collect all best candidates of $F$ using the data-driven method to build the seed lexicon, then our G2P model will be trained on this lexicon and predict possible pronunciations for any foreign word.

%% 4.1 Seed Lexicon  %%
\subsection{Building the Seed Lexicon}
\label{sec4.1}
    The seed lexicon plays an important role in training G2P model.
    In traditional work, the  to the G2P model is an expert-knowledge based seed lexicon, but the seed lexicon in our method consists of all foreign words' best decoding results through  the data-driven method, and the trained G2P model will possess the mapping rules in pronouncing. 
    This paper extracts all foreign words’ audio segments in the CS corpus as driving materials, and then obtain all foreign words' n-best candidates through phonetic decoding and selecting methods, finally all candidates are gathered into a seed lexicon used as  to G2P model. 

%% 4.2 Transformer-based G2P Model  %%
\subsection{Transformer-based G2P Model}
\label{sec4.2}
    Many approaches in sequence modeling and transduction problems generally adopt recurrent neural networks \citep{Kleenankandy2020,Karevan2020,Jiang2020,Gao2020}, but transformer \citep{Vaswani2017} model architecture discards the recurrence module and instead relies entirely on the attention mechanism to draw global dependencies between input and output, which has successfully achieved better results in natural language processing (NLP) tasks \citep{Huang2020,Chen2020,jos2020} such as Neural Machine Translation. In the encoder-decoder architecture of the transformer, the multi-head attention mechanism is widely used so that the model can learn relevant information in different representation sub-spaces. Multi-head attention projects vectors Q (query), K (key), V (value) through $h$ different linear transformations, and then the different attention results are concatenated together: 
    \begin{equation}
        \operatorname{MultiHead}(Q,K,V) = \operatorname{Concat}(head_{1},\cdots,head_{h})W^{O} \notag
    \end{equation}
    where each head is represented: 
    \begin{equation}
        head_{i} = \operatorname{Attention}(QW_{i}^{Q},KW_{i}^{K},VW_{i}^{V}) \notag
    \end{equation}
    the attention approach adopts the scaled dot-product:
    \begin{equation}
        \operatorname{Attention}(Q,K,V) = \operatorname{softmax} \left( \frac{QK^{T}}{\sqrt{d_{k}}} \right) V \notag
    \end{equation}
   In the above formulas, $softmax$ is a activation function used to calculate attention weight, $W$ represents the weight matrix and $d_{k}$ represents the output feature vector dimension. In our work, we use the transformer based G2P model to translate the foreign word from written-form to pronunciation representation in the native language phoneme set, which model is different from the seq2seq-based G2P model proposed in \citep{Huang2019,Rao2015}.

%% 4.3 Improving Seed Lexicon  %%
\subsection{Improving the Seed Lexicon}
\label{sec4.3}
%% summary %%
    In the data-driven processing, the amount of the data used for driving often determines the quality of the results, and more driving materials often have more generalized and diverse results. In this paper, the amount of foreign words' audio segments  will determine whether to generate a more generalized and high-quality pronunciation and solve problems such as the accent diversity in different native speakers. However, in the limited CS corpus, different appearance frequencies cause unequal available materials to different foreign words, which further lead to imbalanced driving processing and different qualities of candidates.

%% 4.3.1 Motivation  %%
\subsubsection{Motivation}
\label{sec4.3.1}
\begin{figure}[ht]
		\centering
		\includegraphics[width=\linewidth]{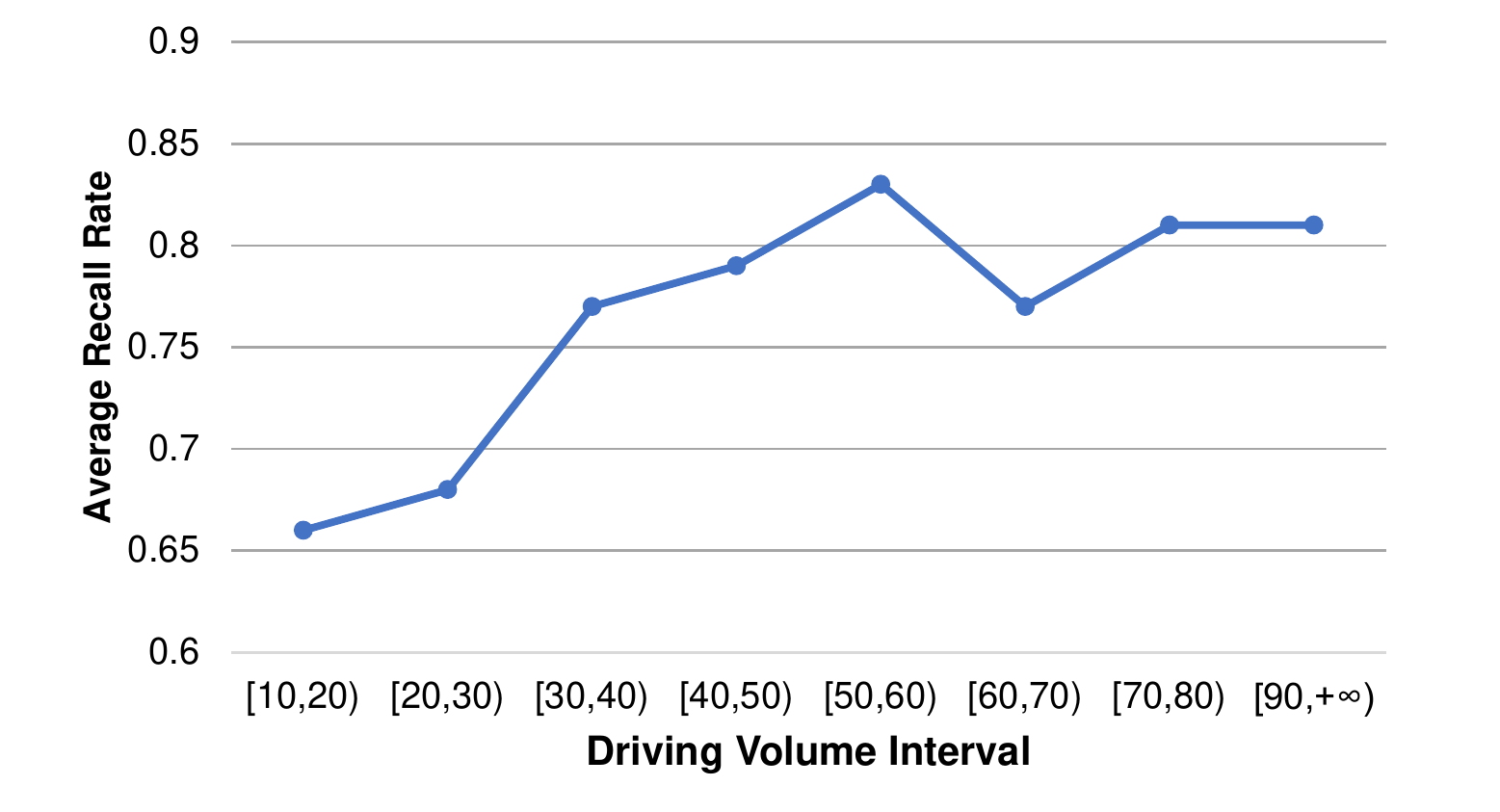}
		\caption{Average recall rate of English words with different driving volume intervals in 5780 Chinese-English code-switching utterances. }
		 \label{issue}
	\end{figure}

    In order to better explain the problem about imbalanced driving materials, we propose a metric ARR (average recall rate), which describes the relationship between recognition accuracy and driving volume (i.e. the number of audio segments) for foreign pronunciation (see \autoref{eq7})
    \begin{equation}
	ARR_{B} =  \frac{1}{N_{B}}\Sigma_{w}\frac{P_{w}}{T_{w}}, \qquad k_{w} \in B
	\label{eq7}
	\end{equation}
	In \autoref{eq7}, $B$ represents a limited driving volume interval, $k_{w}$ represents the driving volume of word $w$, $N_{B}$ denotes the number of foreign words in interval $B$, $P_{w}$ and $T_{w}$ represent the successful recognition count and the true occurrence count of word $w$ respectively, and higher ARR means that better pronunciation quality in CS speech recognition. For example, \autoref{issue} shows the ARR results with  the pronunciations of 880 English words by phonetic decoding and APE method in the 5780 utterances of our Chinese-English intra-sentential CS experiment, and we could observe that the pronunciations generated under small driving volume had poor performance. Such a result  may be attributed to the bottleneck in generating candidates, and unreliable score on a small number of utterances.

%% 4.3.2 Internal Assistance  %%
\subsubsection{Internal Assistance}
\label{sec4.3.2}
    Due to insufficient driving data leading to poor recognition, we hope the pronouncing rules learned with adequate driving materials will provide extra helpful candidates to the scarce data. 
    First, in the raw CS corpus, we define the minimum driving volume limit $\mathcal{K}$ to screen the audio segments of the foreign words whose driving volume is not less than $\mathcal{K}$ as a driving material set $DM_{\mathcal{K}}$, which ensures a more reliable seed lexicon can be obtained. In the limited CS corpus, a larger $\mathcal{K}$ means more reliable driving results in the seed lexicon, but it brings a smaller seed lexicon which is not benefical to training G2P model, so we should weigh the setting of $\mathcal{K}$. 
    
\begin{figure}[ht]
	\centering
    	\includegraphics[width=\linewidth]{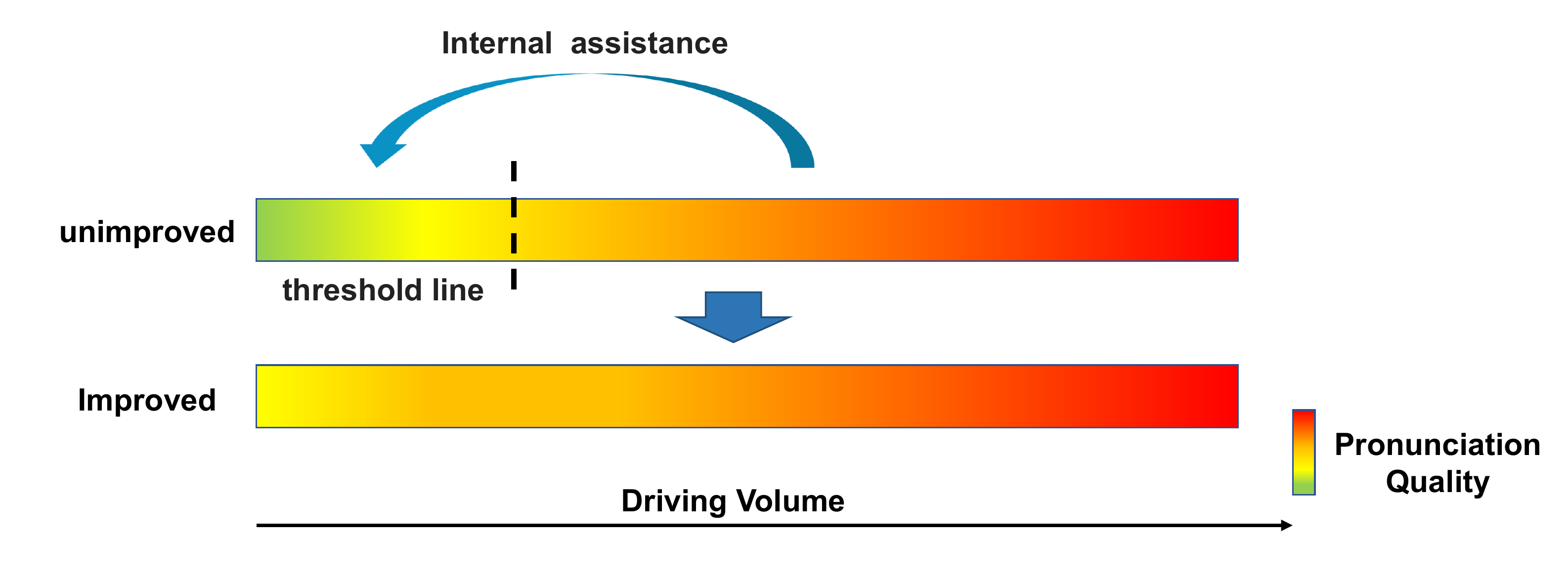}
    	\caption{The proposed internal assistance inspiration is shown on the heat-map of pronunciation quality, where the scarce section will be improved.}
    \label{assistance}
    \end{figure}

\begin{figure}[ht]
	\centering
    	\includegraphics[width=\linewidth]{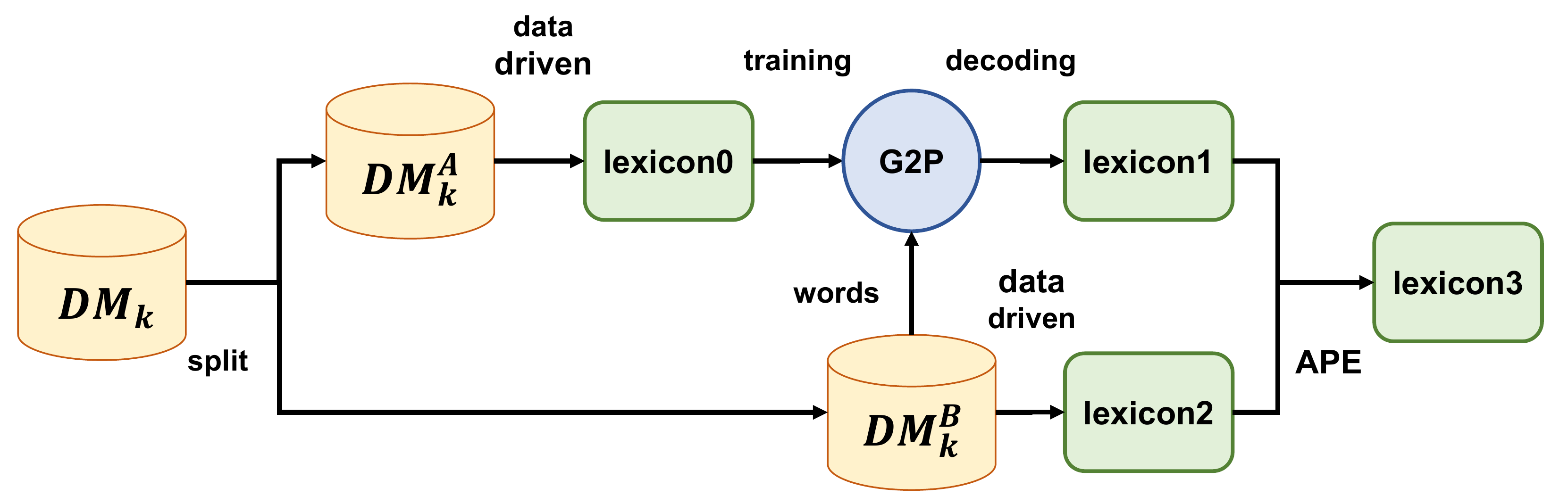}
    	\caption{ The G2P model trained on the sufficient set $DM_{\mathcal{K}}^{A}$ is giving the words in the scarce material set $DM_{\mathcal{K}}^{B}$ extra reference pronunciations }
    \label{assistance_method}
    \end{figure}

    In this work, for the APE based data-driven process, we propose an internal assistance (IA) strategy to improve the pronunciation in scarce materials (see \autoref{assistance}). 
    As \autoref{assistance_method} shows, in the material set $DM_{\mathcal{K}}$ with the restriction  $\mathcal{K}$, we divide $DM_{\mathcal{K}}$ into the sufficient set $DM_{\mathcal{K}}^{A}$ and the scarce set $DM_{\mathcal{K}}^{B}$ by a threshold line (volume) $\mathcal{P}$  we pre-defined ($\mathcal{P}>\mathcal{K}$), and then use the data-driven results \textsl{lexicon0}  of $DM_{\mathcal{K}}^{A}$  to train a G2P model to learn high-quality pronunciation rules, and give foreign words in $DM_{\mathcal{K}}^{B}$ extra reference pronunciations \textsl{lexicon1}.
    Next, we use the APE method to select best candidates between the data-driven results \textsl{lexicon2} of $DM_{\mathcal{K}}^{B}$ and \textsl{lexicon1} to obtain the improved $lexicon3$ for $DM_{\mathcal{K}}^{B}$. Finally, $lexicon0$ and $lexicon3$ were merged and used as a seed lexicon to train our final G2P model.  
    
\section{Experiment}
\label{sec5}
    Our CS experiments were aimed at Chinese and English intra-sentential CS speech recognition, where Chinese and English were native language and foreign language respectively. Assuming that the language model supported Chinese-English mixed grammar, we developed the recognition ability for English words  on the existing Chinese AM, and had compared different seed lexicons generated by different methods.
    
%% 5.1 data %%
\subsection{Data}
\label{sec5.1}
	Our experiments were carried out on the data for ASRU2019 Mandarin-English Code-Switching Challenge provided by Datatang AI Dataset \footnote{www.datatang.ai}, as \autoref{dataset} showed, the speech data spoken by Chinese people using telephone consisted of 500 hours Mandarin training speech data (\textsl{data A}), 200 hours inter/intra-sentential Mandarin-English CS training speech data (\textsl{data B}) and 20 hours inter/intra-sentential Mandarin-English CS testing speech data (\textsl{data C}), additionally the 3gram Chinese-English mixed-language LM was offered also. 
	\begin{table}[ht]
		\centering
		\caption{The information of \textsl{data A}, \textsl{data B}, \textsl{data C} and \textsl{data C*}}
    		\begin{tabular*}{\tblwidth}{@{} CCCCCC@{} }
    			\toprule
    			Data & Type & Hours & Speakers & Utts & English words \\
    			\midrule
    	        \textsl{data A} & Chinese & 500 & 1880 & 561773 & - \\
    	        \textsl{data B} & CS & 200 & 566 & 186479  & 14076\\
    	        \textsl{data C} & CS & 20 & - & 16152 & 6090\\
    	         \textsl{data C*} & CS & 17 & - & 13516 & 5460\\
    			\bottomrule
    		\end{tabular*}%
    \label{dataset}
	\end{table}
	We used \textsl{data A} to train NL-AM, \textsl{data B} to obtain driving materials for foreign words, 
	in order to reduce the impact of the spelling pronunciation of abbreviations, we only considered English words with a length of more than 3 letters and cleaned the \textsl{data C} to obtain a subset \textsl{data C*} \footnote{Many acronyms in the \textsl{data C} were generally short in length, such as UFO, NBA, etc.}, and \textsl{data C*} was used as the final test set.
	Moreover, we built Chinese ASR system using the existing NL-AM and the offered mixed-language LM.

%% 5.2 Preparation %%
\subsection{Preparation Works}
\label{sec5.2}
    This part introduced some preparations about the setting of phoneme sets, the construction and testing of NL-AM, and the preparation of phonetic decoder and mixed-language GMM-HMM in the data-driven.
    
    \paragraph{\textbf{Phoneme-set:}}
    As the phoneme-setting, we chose phonemes from the Chinese lexicon \citep{Bu2018} and the English lexicon \footnote{https://github.com/cmusphinx/cmudict} as AM units, and we shrunk the size of English phonemes-set by merging variants to alleviate the intensity on sparse English-part in CS utterances, which resulted into 214 Chinese phonemes and 39 English phonemes to NL-AM or mixed-language GMM-HMM.
    
    \paragraph{\textbf{NL-AM:}}
	In preparing NL-AM, we split \textsl{data A} into 95\% part to train Chinese NL-AM and 5\% part  to validate Chinese NL-AM, the model was composed of 40-dimensional MFCC (Mel Frequency Cepstral Coefficients) and 3-dimensional pitch input features, 6 layers (625 nodes per layer) Time Delay Neural Network \citep{Peddinti2015}  with LF-MMI \citep{Povey2016} training method and 3944 clustering nodes. As a result, we got an 8.81\% character error rate (CER) on the 5\% part.
	
    \paragraph{\textbf{Phonetic decoder:}}
    Our phonetic decoder was built on the NL-AM and the bi-gram phonetic LM  trained on the alignment results of \textsl{data A}, then we got accented phoneme sequences with high acoustic weight from this decoder.
    
    \paragraph{\textbf{Audio segmentation:}}
    The timestamps for segmentation were obtained through forced-alignment on the Chinese-English mixed-language GMM-HMM model trained on \textsl{data B}, where the model units were a union of Chinese phonemes and English phonemes.
    
    \paragraph{\textbf{LM weight:}}
    We adopted the LM weight using parameter tuning, optimizing the integer value in the interval of 7 to 17 on the validation set.
    
%% 5.3 Seed-lexicon and G2P %%
\subsection{Seed lexicon and G2P model}
\label{sec5.3}
    This part mainly introduces the core works in building the seed lexicon and the configuration of transformer-based G2P model.
    
    \paragraph{\textbf{Driving material:}}
    In building the seed lexicon work, we wanted to collect as high quality and sufficient candidates as possible. As mentioned in \autoref{sec4.3.2}, we considered 10/20/30 $\mathcal{K}$ settings to make the driving material set. The size of foreign words and total driving volume of material sets with different $\mathcal{K}$ were showed in \autoref{material}.
    \begin{table}[htbp]
		\centering
		\caption{The size of foreign words and total driving volume in driving material set with different $\mathcal{K}$ restriction}
		\begin{tabular*}{\tblwidth}{@{} CCCC@{} }
			\toprule
			 Material Set & $\mathcal{K}$ & English words & Total Driving Volume \\
			\midrule
		    $DM_{10}$ & 10 & 2733 & 145897 \\
		    $DM_{20}$ & 20 & 1879 & 133928 \\
		    $DM_{30}$ & 30 & 1369 & 121536 \\
			\bottomrule
		\end{tabular*}
		\label{material}
    \end{table}%
    
\paragraph{\textbf{Phonetic decoding:}}
    First we performed the forced alignment operation on the mixed-language GMM-HMM to obtain the timestamps of foreign words, then cut out segments  and sent these segments to the NL-AM based phonetic decoder. In the decoding process, we set a high acoustic weight in order to pursue the real accent. 
    
\paragraph{\textbf{APE method:}}
    We referred to the interface provided by \citep{Zhang2017} in kaldi-tools \citep{Povey2011} to calculate the posterior probability in the original sentence for each pronunciation. In the calculation process, sentences generated their lattices which saved the alignment information of multiple candidate pronunciations, and then we counted the average posterior probability in all host sentences for a specified pronunciation.  Same as \citep{Zhang2017}, we filtered out low-quality pronunciation through the first APE method and then re-generate lattices to re-calculate the posterior probability, and this pruning helped to improve the accuracy of the posterior probability.

\paragraph{\textbf{PCN method:}}
    In the construction of the PCN, we used the rover method in SCTK-tool \citep{Fiscus1997} to align multiple candidates, and then counted the voting information on each edge to calculate the best paths with the highest scores as the best candidates.

\paragraph{\textbf{G2P model:}}
    Our transformer-based g2p model used  3-layer encoders/decoders and 4-head attentions by building on the tensor2tensor library \citep{Vaswani2018}.

%% 5.4 Metric %%
\subsection{Metric}
\label{sec5.4}
\begin{figure}[ht]
	\centering
    	\includegraphics[width=\linewidth]{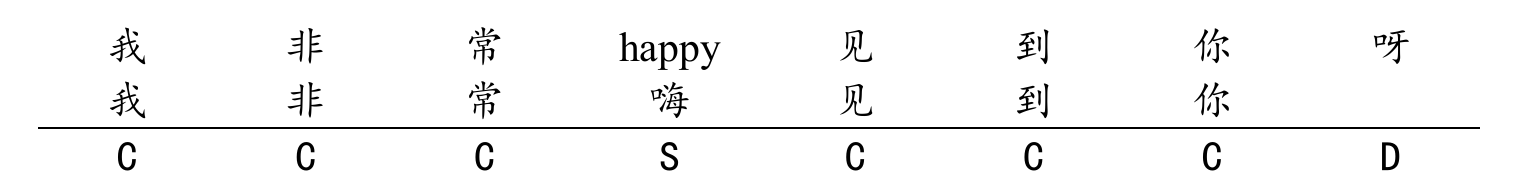}
    	\caption{The detailed alignment example in intra-sentential CS, where line 1 was reference, line 2 was prediction and line3 was alignment types.}
    \label{alignment}
    \end{figure}

	We adopt the mixed error rate (MER) to evaluate the recognition accuracy in CS scenario, where the accuracy of Chinese recognition was calculated by characters, and the accuracy of English recognition was calculated by words. In addition, we used the recall rate to approximate the Chinese character error rate (CER) and the English word error rate (WER). For example, as \autoref{alignment} showed, there were four types of alignment: C:correct, S:substituting, I:Inserting and D:deleting, we could calculate: MER:2/8=25.0\%, CER (Chinese):1-6/7=14.3\%, WER (English):1-0/1=100.0\%.
	
%% 5.5 seed lexicon setting %%
\subsection{Experimental Variables}
\label{sec5.5}
    We considered and obtained different seed lexicons with the following setting:
    \begin{itemize}
        \item \textbf{material set $DM_{\mathcal{K}}$:} specifying the minimum driving volume limit $\mathcal{K}$.
        \item \textbf{selecting method:}  APE or PCN method.
        \item \textbf{internal assistance (IA):} whether to use IA method in seed lexicon with threshold line $\mathcal{P}$.
    \end{itemize}
    All seed lexicons in our experiment trained their G2P models under the same training configurations, then these G2P models gave 4 candidate pronunciations for each English word in the test set. The pronunciation model we adopted  the all-one pronunciation probability for all words.
    
%% 5.6 seed lexicon setting %%
\subsection{Results}
\label{sec5.6}
    We have  completed the following experiments on the 17 hours ASRU2019 Chinese-English CS test-set (\textsl{data C*}). 
    First, as \autoref{exp0} showed, we sent our CS test-set into Chinese ASR system with Chinese lexicon loaded only and got results: 21.5\% CER in Chinese part, 100.0\% WER in English part, and 29.15\% MER overall. Surely the Chinese ASR system couldn't support English words recognition and high WER (for English) caused high CER (for Chinese) because of more insertion errors occurring in the Chinese parts.
    \begin{table}[ht]
		\centering
		\caption{The Chinese-English CS test-set results with only Chinese lexicon loaded }
	   		\begin{tabular}{lll}
	   			\toprule
    			CH(\%) & EN(\%) & MIX(\%) \\
	   			\midrule
          		21.5 & 100.0 & 29.15 \\
	   			\bottomrule
	   		\end{tabular}
    	\label{exp0}
	\end{table}
	
	\begin{table*}[htbp]
		\centering
		\caption{The Chinese-English CS test-set (\textsl{data C*}) results with added pronunciations learned from different seed lexicons}
    		\begin{tabular*}{\tblwidth}{ CLCRRR }
    			\toprule
    		Material Set & Selecting Method & IA & CH(\%) & EN(\%) & MIX(\%) \\
    			\midrule
    			$DM_{30}$ & APE & - & 10.53	& 50.47 & 14.58 \\
    			$DM_{20}$ & APE & - & 9.66 & 43.92 & 13.14 \\
    			$DM_{10}$ & APE (baseline) & - & 9.12 & 38.75 & \textbf{12.13} \\
    			\midrule
    			
    			$DM_{10}$ & APE & $\mathcal{P}=10$ & 9.05 & 38.34 & 12.02 \\
    			$DM_{10}$ & APE & $\mathcal{P}=20$ & 9.00 & 37.28 & 11.87 \\
    			\midrule
    			
    			$DM_{10}$ & PCN$_{5}$ & - & 12.24 & 63.83 & 17.48 \\
    			$DM_{10}$ & PCN$_{10}$ & - & 10.96 & 53.99 & 15.33 \\
    			$DM_{10}$ & PCN$_{20}$ & - & 9.91 & 44.81 & 13.45 \\
    			\midrule
    			
    			$DM_{10}$ & PCN$_{20}$ + APE & - & 8.60	& 34.16 & \textbf{11.19} \\
    			$DM_{10}$ & PCN$_{20}$ + APE &  $\mathcal{P}=20$ & 8.55 & 34.14 & \textbf{11.14} \\
    			
    			\bottomrule
    		\end{tabular*}
    		\label{exp1,2,3,4,5}
	\end{table*}
	
%% exp1: APE method %%
    Next, as \autoref{exp1,2,3,4,5} showed, we made the leap from zero to one on foreign words' recognition, we added predicted pronunciations of all English words in test-set  from G2P models trained on different seed lexicons into the lexicon of Chinese ASR system. 
    In the data-driven process, though the larger the restriction $\mathcal{K}$, the more reliable pronunciations in seed lexicon could be, but the smaller seed lexicon was not good to train the G2P model, the results showed that the larger driving material set achieved better results, where we trained the better G2P model on the seed lexicon  obtained by phonetic decoding and APE method in materials $DM_{10}$, the pronunciation of English words in the test-set predicted by this G2P  achieved 38.75\% WER in English and 12.13\% MER overall, this result we deemed as a baseline.
    
%% exp2: APE + IA method %%
    For the imbalanced driving material issue mentioned in \autoref{sec4.3.1}, we defined two threshold lines $\mathcal{P}=20$ and $\mathcal{P}=30$ in $DM_{10}$, and we compared the assistance of $DM_{10}^{A}(\mathcal{P}=20)$ to $DM_{10}^{B}(\mathcal{P}=20)$  and the assistance of $DM_{10}^{A}(\mathcal{P}=30)$ to $DM_{10}^{B}(\mathcal{P}=30)$. The results showed that this  internal assistance (IA) method had substantial improvement on the quality of the seed lexicon compared to the baseline. The smaller $\mathcal{P}$ has achieved better results and the MER has dropped from 12.13\% to 11.87\% with $\mathcal{P}=20$.
	
%% exp3: PCN method %%
     We then tried to use PCN instead of the APE method to select good candidates. Since there were too many initial phonetic decoding results for each foreign word and that were to greatly increase the computation in batch decoding on PCN, so we had used the APE method to screen a small number of high-quality candidates. In our experiment, $PCN_{n}$ represented a confusion network built on $n$ candidates, and then we used the majority voting strategy on the confusion network to obtain the best summary results. We compared the summary results from the confusion network constructed on 5/10/20 candidates, and experiments showed that the confusion network constructed on more candidates was to generate better summary results, where $PCN_{20}$ got 13.45\% MER.
     However, new variants appeared but original candidates disappeared in the summary results might not perform better, on the whole, it was still slightly worse than the APE method when we directly used the PCN method to build the seed lexicon.
     
%% exp4: PCN + APE method %%
     Since some new variants in PCN summary results might be good candidates, we kept the original candidates and performed the posterior estimation again between the new variants and the original candidates. The results showed that the PCN/APE hybrid method could further improve the quality of the seed lexicon based on the baseline, and the $PCN_{20}$ and APE hybrid method reached 11.19\% MER. 
     
%% exp4: PCN + APE + IA method %%
     Finally, on the basis of the PCN/APE hybrid method, we further added the internal assistance method and the experimental results showed that the recognition result was further improved slightly and reached the best MER 11.14\%. In summary, the seed lexicon obtained by the PCN/APE/IA hybrid method achieved the best recognition in the intra-sentential CS scene.

\section{Conclusion}
\label{sec6}
	%% conclusion
	In this paper, we explore a shortcut to develop intra-sentential code-switching speech recognition in existing NL-AM, where the core work is to obtain the real accented pronunciation which is mapped into  a native language phoneme sequence. To directly build a mixed-language acoustic model for the CS task with a limited CS corpus may be costly and 
	limited performance, instead, we make use of the limited CS corpus and take it as driving material to generate foreign words' pronunciation to meet intra-sentential CS speech recognition task under the NL-AM. 
	
	We propose a data-driven method to create a seed lexicon for training the G2P model to predict the possible pronunciations of any foreign word, where the pronunciation quality of the seed lexicon determines the reliability of prediction. In our experiment we utilize the phonetic decoding method as the source to produce candidates  and compare the different selection methods to screen good candidates.
	Phonetic decoding as a data-driven approach tends to obtain candidates with high acoustic similarity and does well in  generating the native accented pronunciations, and the decoding results from the decoder are often rich and diverse, but contain a lot of noise.
	Hence, for the following screening work, we try the acoustic-based average posterior estimation (APE) \citep{Zhang2017} method and the summary-based phoneme confusion network (PCN) \citep{Huang2019} method to give n-best candidates. It is worth noting that the ROVER-like \citep{Fiscus1997}  PCN method may produce new variants and also at the same time lose the original candidates from the phonetic decoder, so we adopt the PCN/APE hybrid approach to further improve seed lexicon.
	
	However, for different foreign words in the limited CS corpus, the imbalanced driving materials lead to imbalanced driving processing. We therefore developed an internal assistance approach which is to give extra reference candidates to the scarce materials by learning the pronunciation rules in the sufficient materials, where the division of the scarce/sufficient set is determined by the threshold line (i.e. a specified driving volume). Experiments have shown that this internal assistance will further improve the pronunciation quality of the seed lexicon.
	
	Finally, we continue to add internal assistance trick to PCN/APE hybrid method, and experiments show that the PCN/APE/IA hybrid method has a further slight improvement and achieves the best recognition error rate in the intra-sentential CS scenario.
	In fact, in addition to phonetic decoding, PCN and internal assistance method can provide additional meaningful variants, some of them may perform better in the recognition task.
	
	%% feture work
	There are still many unexplored ways to improve the seed lexicon. Under the limited CS corpus, the source to generate candidates for seed lexicon in this article mainly comes from the phonetic decoder, and some new variants generated by PCN or IA method may bring better performance.
	 In future work, we will focus on searching more sources for generating foreign words' pronunciation to improve the pronunciation quality of seed lexicon.

\bibliographystyle{cas-model2-names}
\bibliography{mybib}

\begin{thebibliography}{45}
\expandafter\ifx\csname natexlab\endcsname\relax\def\natexlab#1{#1}\fi
\providecommand{\url}[1]{\texttt{#1}}
\providecommand{\href}[2]{#2}
\providecommand{\path}[1]{#1}
\providecommand{\DOIprefix}{doi:}
\providecommand{\ArXivprefix}{arXiv:}
\providecommand{\URLprefix}{URL: }
\providecommand{\Pubmedprefix}{pmid:}
\providecommand{\doi}[1]{\href{http://dx.doi.org/#1}{\path{#1}}}
\providecommand{\Pubmed}[1]{\href{pmid:#1}{\path{#1}}}
\providecommand{\bibinfo}[2]{#2}
\ifx\xfnm\relax \def\xfnm[#1]{\unskip,\space#1}\fi
%Type = Inproceedings
\bibitem[{Bellegarda(2005)}]{Bellegarda2005}
\bibinfo{author}{Bellegarda, J.R.}, \bibinfo{year}{2005}.
\newblock \bibinfo{title}{Unsupervised, language-independent
  grapheme-to-phoneme conversion by latent analogy},
  \bibinfo{publisher}{North-Holland}. pp. \bibinfo{pages}{140--152}.
\newblock \DOIprefix\doi{10.1016/j.specom.2005.03.002}.
%Type = Article
\bibitem[{Bhuvanagiri and Kopparapu(2010)}]{Bhuvanagiri2010}
\bibinfo{author}{Bhuvanagiri, K.K.}, \bibinfo{author}{Kopparapu, S.K.},
  \bibinfo{year}{2010}.
\newblock \bibinfo{title}{An approach to mixed language automatic speech
  recognition}.
\newblock \bibinfo{journal}{Cocosda} .
%Type = Article
\bibitem[{Bhuvanagirir and Kopparapu(2012)}]{Bhuvanagirir2012}
\bibinfo{author}{Bhuvanagirir, K.}, \bibinfo{author}{Kopparapu, S.K.},
  \bibinfo{year}{2012}.
\newblock \bibinfo{title}{Mixed language speech recognition without explicit
  identification of language}.
\newblock \bibinfo{journal}{American Journal of Signal Processing}
  \bibinfo{volume}{2}.
\newblock \DOIprefix\doi{10.5923/j.ajsp.20120205.02}.
%Type = Article
\bibitem[{Bisani and Ney(2008)}]{Bisani2008}
\bibinfo{author}{Bisani, M.}, \bibinfo{author}{Ney, H.}, \bibinfo{year}{2008}.
\newblock \bibinfo{title}{Joint-sequence models for grapheme-to-phoneme
  conversion}.
\newblock \bibinfo{journal}{Speech Communication} \bibinfo{volume}{50},
  \bibinfo{pages}{434--451}.
\newblock \DOIprefix\doi{10.1016/j.specom.2008.01.002}.
%Type = Inproceedings
\bibitem[{Bu et~al.(2018)Bu, Du, Na, Wu and Zheng}]{Bu2018}
\bibinfo{author}{Bu, H.}, \bibinfo{author}{Du, J.}, \bibinfo{author}{Na, X.},
  \bibinfo{author}{Wu, B.}, \bibinfo{author}{Zheng, H.}, \bibinfo{year}{2018}.
\newblock \bibinfo{title}{Aishell-1: An open-source mandarin speech corpus and
  a speech recognition baseline}.
\newblock \DOIprefix\doi{10.1109/ICSDA.2017.8384449}.
%Type = Inproceedings
\bibitem[{Chan et~al.(2005)Chan, Ching and Lee}]{Chan2005}
\bibinfo{author}{Chan, J.Y.}, \bibinfo{author}{Ching, P.C.},
  \bibinfo{author}{Lee, T.}, \bibinfo{year}{2005}.
\newblock \bibinfo{title}{Development of a cantonese-english code-mixing speech
  corpus}.
%Type = Inproceedings
\bibitem[{Chen et~al.(2016a)Chen, Povey and Khudanpur}]{Chen2016}
\bibinfo{author}{Chen, G.}, \bibinfo{author}{Povey, D.},
  \bibinfo{author}{Khudanpur, S.}, \bibinfo{year}{2016}a.
\newblock \bibinfo{title}{Acoustic data-driven pronunciation lexicon generation
  for logographic languages}.
\newblock \DOIprefix\doi{10.1109/ICASSP.2016.7472699}.
%Type = Inproceedings
\bibitem[{Chen et~al.(2016b)Chen, Pan, Zhao and Yan}]{Chen2016b}
\bibinfo{author}{Chen, M.}, \bibinfo{author}{Pan, J.}, \bibinfo{author}{Zhao,
  Q.}, \bibinfo{author}{Yan, Y.}, \bibinfo{year}{2016}b.
\newblock \bibinfo{title}{Multi-task learning in deep neural networks for
  mandarin-english code-mixing speech recognition}.
\newblock \DOIprefix\doi{10.1587/transinf.2016SLL0004}.
%Type = Article
\bibitem[{Chen and Li(2020)}]{Chen2020}
\bibinfo{author}{Chen, Y.}, \bibinfo{author}{Li, H.}, \bibinfo{year}{2020}.
\newblock \bibinfo{title}{Dam: Transformer-based relation detection for
  question answering over knowledge base}.
\newblock \bibinfo{journal}{Knowledge-Based Systems} \bibinfo{volume}{201-202}.
\newblock \DOIprefix\doi{10.1016/j.knosys.2020.106077}.
%Type = Inproceedings
\bibitem[{Fiscus(1997)}]{Fiscus1997}
\bibinfo{author}{Fiscus, J.G.}, \bibinfo{year}{1997}.
\newblock \bibinfo{title}{Post-processing system to yield reduced word error
  rates: Recognizer output voting error reduction (rover)}.
\newblock \DOIprefix\doi{10.1109/asru.1997.659110}.
%Type = Article
\bibitem[{Ganji et~al.(2019)Ganji, Dhawan and Sinha}]{Ganji2019}
\bibinfo{author}{Ganji, S.}, \bibinfo{author}{Dhawan, K.},
  \bibinfo{author}{Sinha, R.}, \bibinfo{year}{2019}.
\newblock \bibinfo{title}{Iitg-hingcos corpus: A hinglish code-switching
  database for automatic speech recognition}.
\newblock \bibinfo{journal}{Speech Communication} \bibinfo{volume}{110},
  \bibinfo{pages}{76--89}.
\newblock \DOIprefix\doi{10.1016/j.specom.2019.04.007}.
%Type = Article
\bibitem[{Gao et~al.(2020)Gao, Huang, Zhang, Han, Wang, Zhang and
  Lin}]{Gao2020}
\bibinfo{author}{Gao, S.}, \bibinfo{author}{Huang, Y.}, \bibinfo{author}{Zhang,
  S.}, \bibinfo{author}{Han, J.}, \bibinfo{author}{Wang, G.},
  \bibinfo{author}{Zhang, M.}, \bibinfo{author}{Lin, Q.}, \bibinfo{year}{2020}.
\newblock \bibinfo{title}{Short-term runoff prediction with gru and lstm
  networks without requiring time step optimization during sample generation}.
\newblock \bibinfo{journal}{Journal of Hydrology} \bibinfo{volume}{589},
  \bibinfo{pages}{125188}.
\newblock \DOIprefix\doi{10.1016/j.jhydrol.2020.125188}.
%Type = Article
\bibitem[{Ángel González et~al.(2020)Ángel González, Hurtado and
  Pla}]{jos2020}
\bibinfo{author}{Ángel González, J.}, \bibinfo{author}{Hurtado, L.F.},
  \bibinfo{author}{Pla, F.}, \bibinfo{year}{2020}.
\newblock \bibinfo{title}{Transformer based contextualization of pre-trained
  word embeddings for irony detection in twitter}.
\newblock \bibinfo{journal}{Information Processing and Management}
  \bibinfo{volume}{57}.
\newblock \URLprefix \url{https://doi.org/10.1016/j.ipm.2020.102262},
  \DOIprefix\doi{10.1016/j.ipm.2020.102262}.
%Type = Inproceedings
\bibitem[{Guo et~al.(2018)Guo, Xu, Xie and Chng}]{Guo2018}
\bibinfo{author}{Guo, P.}, \bibinfo{author}{Xu, H.}, \bibinfo{author}{Xie, L.},
  \bibinfo{author}{Chng, E.S.}, \bibinfo{year}{2018}.
\newblock \bibinfo{title}{Study of semi-supervised approaches to improving
  english-mandarin code-switching speech recognition}.
\newblock \DOIprefix\doi{10.21437/Interspeech.2018-1974}.
%Type = Inproceedings
\bibitem[{Huang et~al.(2013)Huang, Li, Yu, Deng and Gong}]{Huang2013}
\bibinfo{author}{Huang, J.T.}, \bibinfo{author}{Li, J.}, \bibinfo{author}{Yu,
  D.}, \bibinfo{author}{Deng, L.}, \bibinfo{author}{Gong, Y.},
  \bibinfo{year}{2013}.
\newblock \bibinfo{title}{Cross-language knowledge transfer using multilingual
  deep neural network with shared hidden layers}.
\newblock \DOIprefix\doi{10.1109/ICASSP.2013.6639081}.
%Type = Article
\bibitem[{Huang et~al.(2020)Huang, Mao, Yang, Zhu and Long}]{Huang2020}
\bibinfo{author}{Huang, W.}, \bibinfo{author}{Mao, Y.}, \bibinfo{author}{Yang,
  Z.}, \bibinfo{author}{Zhu, L.}, \bibinfo{author}{Long, J.},
  \bibinfo{year}{2020}.
\newblock \bibinfo{title}{Relation classification via knowledge graph enhanced
  transformer encoder}.
\newblock \bibinfo{journal}{Knowledge-Based Systems} \bibinfo{volume}{206}.
\newblock \DOIprefix\doi{10.1016/j.knosys.2020.106321}.
%Type = Inproceedings
\bibitem[{Huang et~al.(2019)Huang, Zhuang, Liu, Xiao, Zhang and
  Siniscalchi}]{Huang2019}
\bibinfo{author}{Huang, Z.}, \bibinfo{author}{Zhuang, X.},
  \bibinfo{author}{Liu, D.}, \bibinfo{author}{Xiao, X.},
  \bibinfo{author}{Zhang, Y.}, \bibinfo{author}{Siniscalchi, S.M.},
  \bibinfo{year}{2019}.
\newblock \bibinfo{title}{Exploring retraining-free speech recognition for
  intra-sentential code-switching}.
\newblock \DOIprefix\doi{10.1109/ICASSP.2019.8682478}.
%Type = Article
\bibitem[{Jiang et~al.(2020)Jiang, Sun, Mercaldo and Santone}]{Jiang2020}
\bibinfo{author}{Jiang, L.}, \bibinfo{author}{Sun, X.},
  \bibinfo{author}{Mercaldo, F.}, \bibinfo{author}{Santone, A.},
  \bibinfo{year}{2020}.
\newblock \bibinfo{title}{Decab-lstm: Deep contextualized attentional
  bidirectional lstm for cancer hallmark classification}.
\newblock \bibinfo{journal}{Knowledge-Based Systems} \bibinfo{volume}{210},
  \bibinfo{pages}{106486}.
\newblock \URLprefix
  \url{https://linkinghub.elsevier.com/retrieve/pii/S0950705120306158},
  \DOIprefix\doi{10.1016/j.knosys.2020.106486}.
%Type = Article
\bibitem[{Karevan and Suykens(2020)}]{Karevan2020}
\bibinfo{author}{Karevan, Z.}, \bibinfo{author}{Suykens, J.A.},
  \bibinfo{year}{2020}.
\newblock \bibinfo{title}{Transductive lstm for time-series prediction: An
  application to weather forecasting}.
\newblock \bibinfo{journal}{Neural Networks} \bibinfo{volume}{125},
  \bibinfo{pages}{1--9}.
\newblock \DOIprefix\doi{10.1016/j.neunet.2019.12.030}.
%Type = Article
\bibitem[{Kleenankandy and A(2020)}]{Kleenankandy2020}
\bibinfo{author}{Kleenankandy, J.}, \bibinfo{author}{A, A.N.K.},
  \bibinfo{year}{2020}.
\newblock \bibinfo{title}{An enhanced tree-lstm architecture for sentence
  semantic modeling using typed dependencies}.
\newblock \bibinfo{journal}{Information Processing and Management} ,
  \bibinfo{pages}{102362}\DOIprefix\doi{10.1016/j.ipm.2020.102362}.
%Type = Inproceedings
\bibitem[{Laurent et~al.(2010)Laurent, Meignier, Merlin and
  Deleglise}]{Laurent2010}
\bibinfo{author}{Laurent, A.}, \bibinfo{author}{Meignier, S.},
  \bibinfo{author}{Merlin, T.}, \bibinfo{author}{Deleglise, P.},
  \bibinfo{year}{2010}.
\newblock \bibinfo{title}{Acoustics-based phonetic transcription method for
  proper nouns}.
%Type = Inproceedings
\bibitem[{Li et~al.(2011)Li, Fung, Xu and Liu}]{Li2011}
\bibinfo{author}{Li, Y.}, \bibinfo{author}{Fung, P.}, \bibinfo{author}{Xu, P.},
  \bibinfo{author}{Liu, Y.}, \bibinfo{year}{2011}.
\newblock \bibinfo{title}{Asymmetric acoustic modeling of mixed language
  speech}.
\newblock \DOIprefix\doi{10.1109/ICASSP.2011.5947480}.
%Type = Inproceedings
\bibitem[{Li et~al.(2012)Li, Yu and Fung}]{Li2012}
\bibinfo{author}{Li, Y.}, \bibinfo{author}{Yu, Y.}, \bibinfo{author}{Fung, P.},
  \bibinfo{year}{2012}.
\newblock \bibinfo{title}{A mandarin-english code-switching corpus}.
%Type = Inproceedings
\bibitem[{Lin et~al.(2009)Lin, Deng, Yu, Gong, Acero and Lee}]{Lin2009}
\bibinfo{author}{Lin, H.}, \bibinfo{author}{Deng, L.}, \bibinfo{author}{Yu,
  D.}, \bibinfo{author}{Gong, Y.F.}, \bibinfo{author}{Acero, A.},
  \bibinfo{author}{Lee, C.H.}, \bibinfo{year}{2009}.
\newblock \bibinfo{title}{A study on multilingual acoustic modeling for large
  vocabulary asr}.
\newblock \DOIprefix\doi{10.1109/ICASSP.2009.4960588}.
%Type = Inproceedings
\bibitem[{Lu et~al.(2013)Lu, Ghoshal and Renals}]{Lu2013}
\bibinfo{author}{Lu, L.}, \bibinfo{author}{Ghoshal, A.},
  \bibinfo{author}{Renals, S.}, \bibinfo{year}{2013}.
\newblock \bibinfo{title}{Acoustic data-driven pronunciation lexicon for large
  vocabulary speech recognition}.
\newblock \DOIprefix\doi{10.1109/ASRU.2013.6707759}.
%Type = Inproceedings
\bibitem[{Lyu et~al.(2006)Lyu, Lyu, Chiang and Hsu}]{Lyu2006}
\bibinfo{author}{Lyu, D.C.}, \bibinfo{author}{Lyu, R.Y.},
  \bibinfo{author}{Chiang, Y.C.}, \bibinfo{author}{Hsu, C.N.},
  \bibinfo{year}{2006}.
\newblock \bibinfo{title}{Speech recognition on code-switching among the
  chinese dialects}.
\newblock \DOIprefix\doi{10.1109/icassp.2006.1660218}.
%Type = Inproceedings
\bibitem[{Lyu et~al.(2010)Lyu, Tan, Chng and Li}]{Lyu2010}
\bibinfo{author}{Lyu, D.C.}, \bibinfo{author}{Tan, T.P.},
  \bibinfo{author}{Chng, E.S.}, \bibinfo{author}{Li, H.}, \bibinfo{year}{2010}.
\newblock \bibinfo{title}{Seame: A mandarin-english code-switching speech
  corpus in south-east asia}.
%Type = Article
\bibitem[{McGraw et~al.(2013)McGraw, Badr and Glass}]{McGraw2013}
\bibinfo{author}{McGraw, I.}, \bibinfo{author}{Badr, I.},
  \bibinfo{author}{Glass, J.R.}, \bibinfo{year}{2013}.
\newblock \bibinfo{title}{Learning lexicons from speech using a pronunciation
  mixture model}.
\newblock \bibinfo{journal}{IEEE Transactions on Audio, Speech and Language
  Processing} \bibinfo{volume}{21}.
\newblock \DOIprefix\doi{10.1109/TASL.2012.2226158}.
%Type = Inproceedings
\bibitem[{Mendes et~al.(2019)Mendes, Abad, Neto and Trancoso}]{Mendes2019}
\bibinfo{author}{Mendes, C.}, \bibinfo{author}{Abad, A.},
  \bibinfo{author}{Neto, J.P.}, \bibinfo{author}{Trancoso, I.},
  \bibinfo{year}{2019}.
\newblock \bibinfo{title}{Recognition of latin american spanish using
  multi-task learning}.
\newblock \DOIprefix\doi{10.21437/Interspeech.2019-2772}.
%Type = Article
\bibitem[{Modipa et~al.(2013)Modipa, Davel and de~Wet}]{Modipa2013}
\bibinfo{author}{Modipa, T.I.}, \bibinfo{author}{Davel, M.H.},
  \bibinfo{author}{de~Wet, F.}, \bibinfo{year}{2013}.
\newblock \bibinfo{title}{Implications of sepedi/english code switching for asr
  systems}.
\newblock \bibinfo{journal}{Proceedings of the 24th Annual Symposium of the
  Pattern Recognition Association of South Africa (PRASA)} .
%Type = Inproceedings
\bibitem[{Peddinti et~al.(2015)Peddinti, Povey and Khudanpur}]{Peddinti2015}
\bibinfo{author}{Peddinti, V.}, \bibinfo{author}{Povey, D.},
  \bibinfo{author}{Khudanpur, S.}, \bibinfo{year}{2015}.
\newblock \bibinfo{title}{A time delay neural network architecture for
  efficient modeling of long temporal contexts}.
%Type = Inproceedings
\bibitem[{Povey et~al.(2011)Povey, Ghoshal, Boulianne, Burget, Glembek, Goel,
  Hannemann, Motlıcek, Qian, Schwarz, Silovsky, Stemmer and
  Vesely}]{Povey2011}
\bibinfo{author}{Povey, D.}, \bibinfo{author}{Ghoshal, A.},
  \bibinfo{author}{Boulianne, G.}, \bibinfo{author}{Burget, L.},
  \bibinfo{author}{Glembek, O.}, \bibinfo{author}{Goel, N.},
  \bibinfo{author}{Hannemann, M.}, \bibinfo{author}{Motlıcek, P.},
  \bibinfo{author}{Qian, Y.}, \bibinfo{author}{Schwarz, P.},
  \bibinfo{author}{Silovsky, J.}, \bibinfo{author}{Stemmer, G.},
  \bibinfo{author}{Vesely, K.}, \bibinfo{year}{2011}.
\newblock \bibinfo{title}{The kaldi speech recognition toolkit}.
\newblock \DOIprefix\doi{10.1017/CBO9781107415324.004}.
%Type = Inproceedings
\bibitem[{Povey et~al.(2016)Povey, Peddinti, Galvez, Ghahremani, Manohar, Na,
  Wang and Khudanpur}]{Povey2016}
\bibinfo{author}{Povey, D.}, \bibinfo{author}{Peddinti, V.},
  \bibinfo{author}{Galvez, D.}, \bibinfo{author}{Ghahremani, P.},
  \bibinfo{author}{Manohar, V.}, \bibinfo{author}{Na, X.},
  \bibinfo{author}{Wang, Y.}, \bibinfo{author}{Khudanpur, S.},
  \bibinfo{year}{2016}.
\newblock \bibinfo{title}{Purely sequence-trained neural networks for asr based
  on lattice-free mmi}.
\newblock \DOIprefix\doi{10.21437/Interspeech.2016-595}.
%Type = Inproceedings
\bibitem[{Rao et~al.(2015)Rao, Peng, Sak and Beaufays}]{Rao2015}
\bibinfo{author}{Rao, K.}, \bibinfo{author}{Peng, F.}, \bibinfo{author}{Sak,
  H.}, \bibinfo{author}{Beaufays, F.}, \bibinfo{year}{2015}.
\newblock \bibinfo{title}{Grapheme-to-phoneme conversion using long short-term
  memory recurrent neural networks}.
\newblock \DOIprefix\doi{10.1109/ICASSP.2015.7178767}.
%Type = Article
\bibitem[{Rugchatjaroen et~al.(2019)Rugchatjaroen, Saychum, Kongyoung,
  Chootrakool, Kasuriya and Wutiwiwatchai}]{Rugchatjaroen2019}
\bibinfo{author}{Rugchatjaroen, A.}, \bibinfo{author}{Saychum, S.},
  \bibinfo{author}{Kongyoung, S.}, \bibinfo{author}{Chootrakool, P.},
  \bibinfo{author}{Kasuriya, S.}, \bibinfo{author}{Wutiwiwatchai, C.},
  \bibinfo{year}{2019}.
\newblock \bibinfo{title}{Efficient two-stage processing for joint sequence
  model-based thai grapheme-to-phoneme conversion}.
\newblock \bibinfo{journal}{Speech Communication} \bibinfo{volume}{106},
  \bibinfo{pages}{105--111}.
\newblock \URLprefix \url{https://doi.org/10.1016/j.specom.2018.12.003},
  \DOIprefix\doi{10.1016/j.specom.2018.12.003}.
%Type = Article
\bibitem[{Sankoff and Poplack(1981)}]{Sankoff1981}
\bibinfo{author}{Sankoff, D.}, \bibinfo{author}{Poplack, S.},
  \bibinfo{year}{1981}.
\newblock \bibinfo{title}{A formal grammar for code‐switching 1}.
\newblock \bibinfo{journal}{Paper in Linguistics} \bibinfo{volume}{14}.
\newblock \DOIprefix\doi{10.1080/08351818109370523}.
%Type = Inproceedings
\bibitem[{Shen et~al.(2011)Shen, Wu, Yang and Hsu}]{Shen2011}
\bibinfo{author}{Shen, H.P.}, \bibinfo{author}{Wu, C.H.},
  \bibinfo{author}{Yang, Y.T.}, \bibinfo{author}{Hsu, C.S.},
  \bibinfo{year}{2011}.
\newblock \bibinfo{title}{Cecos: A chinese-english code-switching speech
  database}.
\newblock \DOIprefix\doi{10.1109/ICSDA.2011.6085992}.
%Type = Inproceedings
\bibitem[{Sivasankaran et~al.(2019)Sivasankaran, Srivastava, Sitaram, Bali and
  Choudhury}]{Sivasankaran2019}
\bibinfo{author}{Sivasankaran, S.}, \bibinfo{author}{Srivastava, B.M.L.},
  \bibinfo{author}{Sitaram, S.}, \bibinfo{author}{Bali, K.},
  \bibinfo{author}{Choudhury, M.}, \bibinfo{year}{2019}.
\newblock \bibinfo{title}{Phone merging for code-switched speech recognition}.
\newblock \DOIprefix\doi{10.18653/v1/w18-3202}.
%Type = Article
\bibitem[{Smith(2000)}]{Smith2000}
\bibinfo{author}{Smith, C.L.}, \bibinfo{year}{2000}.
\newblock \bibinfo{title}{Handbook of the international phonetic association: A
  guide to the use of the international phonetic alphabet (1999)}.
\newblock \bibinfo{journal}{Phonology} \bibinfo{volume}{17}.
\newblock \DOIprefix\doi{10.1017/S0952675700003894}.
%Type = Inproceedings
\bibitem[{Tsujioka et~al.(2016)Tsujioka, Sakti, Yoshino, Neubig and
  Nakamura}]{Tsujioka2016}
\bibinfo{author}{Tsujioka, S.}, \bibinfo{author}{Sakti, S.},
  \bibinfo{author}{Yoshino, K.}, \bibinfo{author}{Neubig, G.},
  \bibinfo{author}{Nakamura, S.}, \bibinfo{year}{2016}.
\newblock \bibinfo{title}{Unsupervised joint estimation of grapheme-to-phoneme
  conversion systems and acoustic model adaptation for non-native speech
  recognition}.
\newblock \DOIprefix\doi{10.21437/Interspeech.2016-919}.
%Type = Inproceedings
\bibitem[{Vaswani et~al.(2018)Vaswani, Bengio, Brevdo, Chollet, Gomez, Gouws,
  Jones, Łukasz Kaiser, Kalchbrenner, Parmar, Sepassi, Shazeer and
  Uszkoreit}]{Vaswani2018}
\bibinfo{author}{Vaswani, A.}, \bibinfo{author}{Bengio, S.},
  \bibinfo{author}{Brevdo, E.}, \bibinfo{author}{Chollet, F.},
  \bibinfo{author}{Gomez, A.N.}, \bibinfo{author}{Gouws, S.},
  \bibinfo{author}{Jones, L.}, \bibinfo{author}{Łukasz Kaiser},
  \bibinfo{author}{Kalchbrenner, N.}, \bibinfo{author}{Parmar, N.},
  \bibinfo{author}{Sepassi, R.}, \bibinfo{author}{Shazeer, N.},
  \bibinfo{author}{Uszkoreit, J.}, \bibinfo{year}{2018}.
\newblock \bibinfo{title}{Tensor2tensor for neural machine translation}.
%Type = Inproceedings
\bibitem[{Vaswani et~al.(2017)Vaswani, Shazeer, Parmar, Uszkoreit, Jones,
  Gomez, Łukasz Kaiser and Polosukhin}]{Vaswani2017}
\bibinfo{author}{Vaswani, A.}, \bibinfo{author}{Shazeer, N.},
  \bibinfo{author}{Parmar, N.}, \bibinfo{author}{Uszkoreit, J.},
  \bibinfo{author}{Jones, L.}, \bibinfo{author}{Gomez, A.N.},
  \bibinfo{author}{Łukasz Kaiser}, \bibinfo{author}{Polosukhin, I.},
  \bibinfo{year}{2017}.
\newblock \bibinfo{title}{Attention is all you need}.
%Type = Inproceedings
\bibitem[{YIlmaz et~al.(2016)YIlmaz, Heuvel and Leeuwen}]{Yilmaz2016}
\bibinfo{author}{YIlmaz, E.}, \bibinfo{author}{Heuvel, H.V.D.},
  \bibinfo{author}{Leeuwen, D.V.}, \bibinfo{year}{2016}.
\newblock \bibinfo{title}{Investigating bilingual deep neural networks for
  automatic recognition of code-switching frisian speech}.
\newblock \DOIprefix\doi{10.1016/j.procs.2016.04.044}.
%Type = Inproceedings
\bibitem[{Yu et~al.(2009)Yu, Li, Liu, Wu, Gong and Acero}]{Yu2009}
\bibinfo{author}{Yu, D.}, \bibinfo{author}{Li, D.}, \bibinfo{author}{Liu, P.},
  \bibinfo{author}{Wu, J.}, \bibinfo{author}{Gong, Y.}, \bibinfo{author}{Acero,
  A.}, \bibinfo{year}{2009}.
\newblock \bibinfo{title}{Cross-lingual speech recognition under runtime
  resource constraints}.
\newblock \DOIprefix\doi{10.1109/ICASSP.2009.4960553}.
%Type = Inproceedings
\bibitem[{Zhang et~al.(2017)Zhang, Manohar, Povey and Khudanpur}]{Zhang2017}
\bibinfo{author}{Zhang, X.}, \bibinfo{author}{Manohar, V.},
  \bibinfo{author}{Povey, D.}, \bibinfo{author}{Khudanpur, S.},
  \bibinfo{year}{2017}.
\newblock \bibinfo{title}{Acoustic data-driven lexicon learning based on a
  greedy pronunciation selection framework}, \bibinfo{publisher}{International
  Speech Communication Association}. pp. \bibinfo{pages}{2541--2545}.
\newblock \DOIprefix\doi{10.21437/Interspeech.2017-588}.

\end{thebibliography}

\end{document}